\documentclass[12pt]{article}
\usepackage{graphicx}

\newsavebox{\sboxpubnumber}
\newsavebox{\sboxpubdate}
\newcommand{\pubdate}[1]{\begin{lrbox}{\sboxpubdate}{#1}\end{lrbox}}
\newcommand{\pubnumber}[1]{\begin{lrbox}{\sboxpubnumber}{\begin{tabular}{l}
        #1 \\ 
                                 \usebox{\sboxpubdate}
                                 \end{tabular}}
                           \end{lrbox}
                           \pubblock}
\newcommand{\Title}[1]{\begin{center} {\Large #1 } \end{center}}
\newcommand{\Author}[1]{\begin{center}{ \sc #1} \end{center}}
\newcommand{\Address}[1]{\begin{center}{ \it #1} \end{center}}
\newcommand{\andauth}{\begin{center}{and} \end{center}}
\newcommand{\pubblock}{\rightline{
                        \usebox{\sboxpubnumber}}}
\newenvironment{Abstract}{\begin{quotation}  }{\end{quotation}}
\newenvironment{Presented}{\begin{quotation} \begin{center}
             PRESENTED AT\end{center}\bigskip
      \begin{center}\begin{large}}{\end{large}\end{center}
      \end{quotation}}
\newcommand{\Acknowledgements}{\bigskip  \bigskip \begin{center} \begin{large}
             \bf ACKNOWLEDGEMENTS \end{large}\end{center}}
\begin{document}

%%%%%%%%%%%%%%%%%%%%%%%%%%%%%%%%%%%%%%%%%%%%%%%%%%%%%%%%%%%%%%%%%%%%%%%%
%%
%% START EDITING HERE!
%%
%%%%%%%%%%%%%%%%%%%%%%%%%%%%%%%%%%%%%%%%%%%%%%%%%%%%%%%%%%%%%%%%%%%%%%%%
\begin{titlepage}
\pubdate{\today}                    %fill in the date
%\pubnumber{XXX-XXXXX \\ YYY-YYYYYY} %preprint number(s)

\vfill
\Title{Cosmological Vorticity Perturbations,
Gravitomagnetism, and Mach's Principle}
\vfill
\Author{Christoph Schmid}
%\footnote{And possible funding acknowledgements.
%                           DELETE THIS FOOTNOTE IF UNNECESSARY!!}}
\Address{Theoretische Physik, ETH-Z\"urich, CH-8093 Z\"urich\\
Switzerland}

\vfill
%\andauth
%\vfill
%\Author{Your Coauthors Name}
%\Address{Department, Institute \\
%         Postal address}
%\vfill
\begin{Abstract}
  The axes of gyroscopes experimentally define non-rotating frames.
  But what physical cause governs the time-evolution of gyroscope axes
  ?  Starting from an unperturbed, spatially flat FRW cosmology, we
  consider cosmological vorticity perturbations (i.e. vector
  perturbations, rotational perturbations) at the linear level. We
  ask: Will cosmological rotational perturbations drag the axis of a
  gyroscope relative to the directions (geodesics) to galaxies beyond
  the rotational perturbation? We cast the laws of Gravitomagnetism
  into a form showing clearly the close correspondence with the laws
  of ordinary magnetism. Our results are:
\begin{enumerate}
\item[1)] The dragging of a gyroscope axis by rotational perturbations
  beyond the $\dot{H}$ radius ($H$ = Hubble constant) is exponentially
  suppressed.
\item[2)] If the perturbation is a homogeneous rotation inside a
  radius significantly larger than the $\dot{H}$ radius, then the
  dragging of the gyroscope axis by the rotational perturbation is
  exact for any equation of state for cosmological matter.
\item[3)] The time-evolution of a gyroscope axis exactly follows a
  specific average of the matter inside the $\dot{H}$ radius for any
  equation of state.
\end{enumerate} 
In this precise sense Mach's Principle follows from 
cosmology with Einstein Gravity.
\end{Abstract}
\vfill

\begin{Presented}
    COSMO-01 \\
    Rovaniemi, Finland, \\
    August 29 -- September 4, 2001
\end{Presented}
\vfill
\end{titlepage}
%\def\thefootnote{\fnsymbol{footnote}}
%\setcounter{footnote}{0}

%%%%%%%%%%%%%%%%%%%%%%%%%%%%%%%%%%%%%%%%%%%%%%%%%%%%%%%%%%%%%%%%%%%%%%%%
% The document starts here
%%%%%%%%%%%%%%%%%%%%%%%%%%%%%%%%%%%%%%%%%%%%%%%%%%%%%%%%%%%%%%%%%%%%%%%%
\section{Introduction and Conclusions}

The spin axes of gyroscopes define our local nonrotating frame
experimentally. But what physical cause governs the time-evolution of
gyroscope axes?  Newton invoked ``absolute space''. Mach~\cite{1} made
the hypothesis that the nonrotating frame is determined (caused) ``in
some way'' by the motions of all masses in the universe. He did not
know any mechanism, any force, which would have this effect.

We shall show that General Relativity, specifically Gravitomagnetism,
in the context of the Friedmann-Robertson-Walker (FRW) cosmology,
spatially flat, with rotational perturbations (treated as linear
perturbations) makes predictions in total agreement with Mach's
hypothesis as formulated above. 

Thirring in 1918 \cite{2} analyzed the partial dragging of inertial
frames inside a rotating infinitely thin spherical shell of uniform
surface mass density and total mass $M$. He found that general
relativity in the weak field approximation and in first order in the
angular velocity $\Omega_{\rm shell}$ predicts an acceleration of test
particles in the interior corresponding to a Coriolis force, which
could be eliminated by going to a reference frame which is rotating
with $\Omega' = f_{\rm drag} \Omega_{\rm shell}.$ In the weak field
approximation, $G_{N}M \ll R,$ he obtained the dragging fraction
$f_{\rm drag} = \frac{4}{3} \frac{G_{N}M}{R} \ll 1.$ Lense and Thirring
\cite{2} made the corresponding analysis outside a rotating star.

Einstein, after initially being inspired by Mach's ideas, concluded in
1949 \cite{3}: "Mach conjectures that inertia would have to depend
upon the interaction of masses, precisely as was true for Newton's
other forces, a conception which for a long time I considered as in
principle the correct one. It presupposes implicitly, however, that
the basic theory should be of the general type of Newton's mechanics :
masses and their interactions as the original concepts.  The attempt
at such a solution does not fit into a consistent field theory, as
will be immediately recognized."  We shall show in section 4, how this
apparent contradiction with Mach's conjecture is resolved in General
Relativity, specifically in Gravitomagnetism.

Goedel in 1949 \cite{4} presented a model which contradicts Mach's
principle, but since it contains closed time-like curves, it is not
relevant for physics.

Brill and Cohen in 1966 \cite{5} again considered a very thin rotating
spherical shell to lowest order in the rotation frequency, but now to
all orders in the mass $M$ of the shell, i.e. they treated geometry as
a perturbation of Schwarzschild geometry.  They found that space-time
is flat throughout the interior and that the interior Minkowski space
is rotating relative to the asymptotic Minkowski space. For large
masses, whose Schwarzschild radius approaches the shell radius, Brill
and Cohen found that the induced rotation inside approaches the
rotation of the shell, i.e. the dragging of the inertial frame inside
by the rotating matter of the shell becomes perfect.

A cosmologically more relevant example is a rotating ball of dust and
a gyroscope at the center.  In a weak-field approximation with a
Minkowski background this is the same as a superposition of a sequence
of Thirring shells, which gives $ f_{\rm drag}$ of the order of $ G_{N}
M_{\rm ball}/R_{\rm ball}.$ The Friedman equation gives $ G_{N} \rho $ of
order of $ H^2 \equiv R_{H}^{-2}, $ if the curvature term is not
dominant. Hence $ f_{\rm drag} $ is of order of $ R_{\rm ball}^{2}/ R_{H}^{2}
$, valid for $ R_{\rm ball} \ll R_{H}.$ Hence the dragging factor
approaches the order of $ 1 $ only for $ R_{\rm ball} \rightarrow R_{H} $,
but in this case the weak field approximation on a Minkowski
background breaks down.  See also the discussion of Mach's principle
in Misner, Thorne, and Wheeler \cite{6}.

C. Klein \cite{7} analyzed a thin spherical shell with empty interior
embedded in a Friedmann universe. The shell is required to follow the
expanding motion of the surrounding cosmic dust.  Rotations are
treated to first order in the angular velocity. In his treatment the
dragging coefficient tends to one, if nearly "the whole mass of the
universe" is concentrated in the thin shell.

At a conference on Mach's Principle in 1993, edited by Barbour and
Pfister \cite{8}, the question "Is general relativity with appropriate
boundary conditions of closure of some kind perfectly Machian?"  was
put to the participants at the end of the conference.  The vote was
"No" with a clear majority (page 106 of \cite{8}).

In this paper we discuss a realistic cosmological model and use
cosmological perturbation theory including super-horizon perturbations
(instead of perturbations around Minkowski space or around the
Schwarzschild solution), we use realistic cosmological matter (instead
of matter with a contrived energy-momentum-stress tensor), and we
analyze the most general cosmological perturbations in the vorticity
sector (instead of toy models like rigidly rotating thin shells). We
consider a background FRW cosmology, spatially flat, with linear
vorticity perturbations, and we obtain the following specific results:

\begin{enumerate}
\item[1)] The dragging of gyroscope axes by rotational perturbations
        beyond their $\dot{H}$ radius is exponentially suppressed,
        where $H=$ Hubble constant.
      \item[2)] For a homogeneous rotation of cosmological matter
        inside a perturbation radius $R_p$, the dragging of the axis
        of a gyroscope at the center approaches exact dragging
        exponentially fast as $R_p$ increases beyond the $\dot{H}$
        radius. This holds for any equation of state for cosmological
        matter. 
 \item[3)] A gyroscope axis exactly follows a specific average of the
   energy flow in the universe with an exponential cutoff outside the
   $\dot{H}$ radius for any equation of state. 
\end{enumerate}

In this precise sense we have shown that for a spatially flat FRW
Universe with rotational perturbations (added at the linear level)
Mach's Principle on nonrotating frames follows from General Relativity
(without the need to impose any boundary condition of closure).

\section{Vorticity Perturbations}

The vorticity $\vec{\omega}$ of a velocity field $\vec{v}$ is defined
by $\vec{\omega} = {\rm curl}~\vec{v}$.  Every velocity field can be
decomposed into a potential flow, ${\rm curl}~\vec{v} = 0$, and a
vorticity flow, ${\rm div}~\vec{v} = 0$.  The simplest vorticity
perturbation is a spherical region which rotates with a homogeneous
angular velocity $\Omega$ around an axis through its center,
\begin{math} \vec{v} = \Omega  \vec{e} \times \vec{r} \end{math}.
A more realistic cosmological perturbation is one with a Gaussian cutoff,
$\Omega (r) = \Omega_{0} ~{\rm exp}~ (- r^{2}/R_{p}^{2})$. 
In the cosmological context a perturbation in a spherical region 
with a velocity field regular at the origin is much more appropriate 
than the vorticity field familiar from hydrodynamics with
cylindrical symmetry and a vortex-line singularity in the center.

The most general vorticity field $\vec{v}$ can be given by an
expansion in vector spherical harmonics
$\vec{X}_{\ell,m}(\theta,\phi)$, ref.~\cite{9}: 
\begin{equation}
\vec{X}_{\ell, m}(\theta,\phi) \;=\;\vec{L}\;Y_{\ell,m}(\theta, \phi),~~
  \vec{L} = - i\,\vec{r} \times \vec{\nabla}~,
\label{eq:1}
\end{equation}
\begin{equation}
\vec{v}(\vec{r}) \;=\; \sum_{\ell,m} A_{\ell,m} (r)~ 
\vec{X}_{\ell,m} (\theta, \phi) + \sum_{\ell,m} {\rm curl}~ \left[
  B_{\ell,m} (r)~
\vec{X}_{\ell,m} (\theta,\phi)\right]~.
\label{eq:2}
\end{equation}
The simplest case is $\ell=1$, which can be oriented such that $m=0$:
The $A$-solution is a homogeneous rotation at each value of $r$ (see
the simple examples above), which is called a ``toroidal'' field. The
$B$-solution is called a ``poloidal field''. If the velocity field is
toroidal, the vorticity field is poloidal and vice versa. These two
configurations are familiar from ordinary magnetostatics, where ${\rm
  curl}~\vec{B} = 4 \pi \vec{J}$. 

\section{Cosmological Perturbation Theory}

For superhorizon perturbations we need a general relativistic
treatment. This was developed by J. Bardeen in his Phys.~Rev. article
of 1980 (ref.~\cite{10}). In linear perturbation theory one has three
decoupled sectors of perturbations: scalar (density perturbations),
vector (vorticity or rotational perturbations), tensor (gravitational
wave perturbations) under rotations in 3-space. In the vector sector
all perturbation quantities must be built from 3-vector fields with
zero divergence, i.e. from pure vorticity fields.

The perturbation  $\delta g_{00}$ is a scalar under
3-rotations, hence
\begin{equation}
\delta g_{00} = 0~.
\label{eq:3}
\end{equation}
Therefore the slicing of space-time by space-like hypersurfaces of
fixed time $\Sigma_t$ has measured times between slices for $\Delta t
=1$ (lapse) unperturbed. 

In contrast it is not useful (although possible) to insist on a
time-orthogonal foliation (hence time-orthogonal coordinates). Rather
we allow that the lines $\vec{x} = {\rm const}$~ are not orthogonal on
the hypersurfaces of constant time $\Sigma_t$. It is useful to
introduce the concept of a field of fiducial observers (FIDO's), which
are located at fixed values of $\vec{x}$, ref.~\cite{11}. The shift-3-vector
field $\vec{\beta}$ is defined as the 3-velocity $d\vec{v}/dt$ of the
FIDO relative to the normals on $\Sigma_t$.

One can easily show \cite{10} that
it is always possible to find a gauge transformation (coordinate
transformation) such that the new coordinates give
\begin{equation}
\delta^{(3)}\;g_{ij}\;=\;0~.
\label{eq:4}
\end{equation}
Therefore the 3-geometry of $\Sigma_t$ is unperturbed. We take the FRW
background spatially flat, hence $\Sigma_t$ has the geometry of the
Euclidean 3-space. In the gauge of eq.~(\ref{eq:4}) $\Sigma_t$ has
comoving  Cartesian coordinates,
\begin{equation}
ds^2\;=\; - dt^2 + a^2 (t) (dx^i)^2 + 2 a \beta_i\,dx^i\,dt~.
%\label{eq:5}
\end{equation}
Geodesics on $\Sigma_t$ are straight lines on our choice of chart. In
all other gauges the coordinates of Euclidean 3-space are such that if
the coordinates are Cartesian at one time, at all other times the
coordinate lines (e.g. the x-axis) are no longer geodesics, they are
getting wound up relative to geodesics.

Asymptotically (beyond the region of the perturbation) we have FRW
geometry and FRW coordinates, hence distant galaxies (galaxies beyond
the rotational perturbation) are at fixed $\vec{x}$ (comoving
coordinates). Therefore the coordinate axes in our gauge stay fixed
relative to distant galaxies, i.e. the coordinate axes are geodesics
on $\Sigma_t$ and nonrotating relative to distant galaxies. In the
context of rotating black holes such coordinates are called
``star-fixed coordinates''. For brevity of notation we shall employ
this short term, although we always mean ``coordinates fixed to
galaxies beyond the rotational perturbation''.

\section{The Laws of Gravitomagnetism for \\
``star''-fixed FIDO's}

By what dynamical mechanism can cosmological matter in rotational
motion far away influence the spin axes of gyroscopes here? By the
laws of general relativity, more precisely gravitomagnetism. Since we
work to first order in vorticity perturbations, we must analyze
weak-field gravitomagnetism. We cast the laws of weak-field
gravitomagnetism into a form which shows clearly the close
correspondence with the familiar laws of ordinary magnetism
(electromagnetism). We introduce and work with the gravitomagnetic
field $\vec{B}_g$ and the gravitoelectric field $\vec{E}_g$ measured
by fiducial observers (FIDO's) ref.~\cite{11}, where the FIDO is a
crucial concept for our formulation of the laws of gravitomagnetism.
Note that for observers who are free-falling and non-rotating relative
to gyroscopes, i.e. for inertial observers, there are no gravitational
forces, $\vec{E}_g = 0$,~~$\vec{B}_g=0$, by the equivalence principle.

Our choice is to work with ``star''-fixed FIDO's (more precisely
FIDO's fixed by galaxies beyond the rotational perturbation).
``Star''-fixed FIDO's are defined to stay at fixed $\vec{x}$ in the
``star''-fixed coordinate system, and (as in general)
$\bar{e}_{\hat{0}}$(FIDO) $= \bar{u}$(FIDO). Furthermore the spatial
unit vectors $\bar{e}_{\hat{\jmath}}$ of the FIDO's LONB (local
orthonormal basis) point towards fixed distant galaxies.

The 3-momentum measured by FIDO's is denoted by
$\vec{p}=p^{\hat{\jmath}}\; \bar{e}_{\hat{\jmath}}$. Hats on indices
refer to LONB's, the components $p^{\hat{\jmath}}$ are quantities directly
measured by FIDO's. Note that this 3-vector lies in the hyperplane
orthogonal to $\bar{u}$(FIDO), not in the hyperplane $\Sigma_t$. The
time-derivative $d\vec{p}/dt$ measured by FIDO's for a free-falling
test particle gives the operational definition of the gravitational
force,
\begin{equation}
(d\vec{p}/dt)_{\rm free~fall}\;=:\;\vec{F}_g~.
\label{eq:6}
\end{equation}
The gravitoelectric field $\vec{E}_g$ is operationally defined via 
quasistatic test particles,
\begin{equation}
\vec{E}_g \;:=\; \left(\frac{\vec{F}_g}{m}\right)_{\rm
quasistatic~particle} ~.
\label{eq:7}
\end{equation}
$\vec{E}_g=\vec{g}$ is the gravitational acceleration relative to the
FIDO (i.e. measured by a FIDO) for a free-falling particle. Compare
with FIDO's on the surface of the earth (at fixed $r$), who measure
$\vec{g}=-9.8~{\rm m/s}^2~\vec{e}_{\hat{r}}$.  The gravitomagnetic
field $\vec{B}_g$ is defined via the first-order term in the velocity
of the test particle (we put $c=1$),
\begin{equation}
(\vec{F}_g)_{1^{\rm st}~{\rm order~in}~v}\;=\; m (\vec{v} \times \vec{B}_g)~.
\label{eq:8}
\end{equation}

After having given the operational definitions of $\vec{E}_g$ and
$\vec{B}_g$, which are valid for any choice of FIDO's, and having
fixed our choice of FIDO's (``star''-fixed), we can derive the laws of
gravitomagnetism. The results are:

\begin{enumerate}
\item[1)] Equation of motion for test particles. We give it for
  vorticity perturbations of Minkowski space (for simplicity in this
  conference report), 
\begin{equation}
d\vec{p}/dt \;=\; \varepsilon\,\left( \vec{E}_g + \vec{v} \times
  \vec{B}_g\right)~, 
\label{eq:9}
\end{equation}
where $\varepsilon$ is the energy measured by the FIDO. This law is
valid for free-falling particles of arbitrary velocities (e.g.
photons), but only for ``star''-fixed FIDO's. For other FIDO's there
will be terms bilinear in the velocity. Our law should be compared to
the Lorentz law of electromagnetism $d\vec{p}/dt = q (\vec{E} +
\vec{v} \times \vec{B})$.
\item[2)] Equation of motion for the spin of a test particle or
  gyroscope at rest with respect to the FIDO.
\begin{equation}
d\vec{S}/dt\;=\;- \frac 1 2 ~\vec{B}_g \times \vec{S}~.
\label{eq:10}
\end{equation}
This gives an angular velocity of precession of the spin axis relative
to the axes of the FIDO (which in our case is relative to the
direction to distant galaxies), 
\begin{equation}
\vec{\Omega}_{\rm precession} \;=\; -  \frac 1 2~ \vec{B}_g~.
\label{eq:11}
\end{equation}
This should be compared to the analogous precession of the intrinsic
angular momentum (spin) of classical charged particles in an
electromagnetic field, ~$\vec{\Omega}_{\rm precession} = + \frac 1 2
~\frac q m ~\vec{B}.$ Note again that for an inertial observer
$d\vec{p}/dt = 0,~d\vec{S}/dt =0.$ Everything 
depends on the choice of the field of FIDO's.  Using eqs.~(\ref{eq:8})
and (\ref{eq:11}) we obtain $\vec{F}_{g} = 2m(\vec{\Omega}_{\rm
  precession} \times \vec{v}),$ i.e. the Coriolis force, where we must
remember that $\vec{\Omega}_{\rm precession}$ is minus the rotation
velocity of the FIDO relative to the gyroscope axis.  Since we are
working to lowest order in $\vec{\Omega}$ the centrifugal term
vanishes.  A homogeneous gravitomagnetic field can be transformed away
completely by going to a rigidly rotating coordinate system, i.e.
physics on a merry-go-round in Minkowski space is equivalent to
physics in a homogeneous gravitomagnetic field. This is analogous to
the equivalence between physics in a homogeneous Newtonian
gravitational field and physics in a linearly accelerated frame.
\item[3)] Relation between $\vec{E}_g,~\vec{B}_g$ and the shift vector
  $\vec{\beta}$. The gravitational metric perturbations
  $\delta\,g_{\mu\nu}$, i.e. the gravitational potentials for
  vorticity perturbations in ``star''-fixed coordinates  are given by
  the shift vector $\vec{\beta}$, 
  which is connected to $\vec{E}_g$ and $\vec{B}_g$ by
\begin{equation}
\vec{E}_g \;=\; - \partial_t \;\vec{\beta},~~~ 
\vec{B}_g \;=\; \vec{\nabla} \times \vec{\beta}~.
\label{eq:12}
\end{equation}
We see that the shift vector $\vec{\beta}$ plays exactly the same role
for gravitomagnetism as the electromagnetic vector potential $\vec{A}$
for electromagnetism. Therefore we can write $\vec{\beta}=\vec{A}_g$,
and call it the gravitomagnetic vector potential.

\item[4)] Einstein's $G_{\hat{0}\hat{\jmath}}$ equation for weak-field
  gravitomagnetism: 

A) In Minkowski background
\begin{equation}
\nabla \times \vec{B}_g \;=\; - 16\,\pi~G_N~\vec{J}_\varepsilon ~,
\label{eq:13}
\end{equation}
where $\vec{J}_\varepsilon$ is the energy current, and in linear
perturbation theory $\vec{J}_\varepsilon = (\rho + p)\vec{v}$ with
$\vec{v} = \frac{d\vec{x}}{dt}$. Compare Einstein's
$G_{\hat{0}\hat{\jmath}}$ equation to the Amp\`ere-Maxwell equation in
electromagnetism, $\nabla \times \vec{B} - \partial_t\;\vec{E}=+
4\pi\;\vec{J}_q$, where $\vec{J}_q$ is the charge current. In
weak-field gravitomagnetism for ``star''-fixed FIDO's the
$G_{\hat{0}\hat{\jmath}}$ equation is a constraint equation (i.e. no time
derivatives, an equation at one moment of time with the same form as
Amp\`ere's law of magnetostatics), while the Amp\`ere-Maxwell equation
contains a time-derivative and is not a constraint equation.

B) In a FRW background eq.~(\ref{eq:13}) gets an extra term, 
\begin{equation}
\nabla \times \vec{B}_g - 4 \dot{H}\, \vec{\beta} \;=\; -
16\,\pi\;G_N\;\vec{J}_\varepsilon~. 
\label{eq:14}
\end{equation}

>From $\dot{H}=-4\pi G_N (\rho + p)$ we see that $\dot{H} \leq 0$ for
$p\geq -\rho$, therefore we define the~$\dot{H}$ radius by
$R_{\dot{H}}^2 = (-\dot{H})^{-1}$, and we define $\mu^2=-4\dot{H} = 4
R_{\dot{H}}^{-2}$. 
We insert the vector
potential $\vec{A}_g=\vec{\beta}$, we use ~${\rm div}~\vec{A}_g =0$
and ~${\rm curl~curl}~\vec{A}_g = - \Delta~\vec{A}_g$, hence we obtain
by Fourier transformation
\begin{equation}
\left( k^2_{\rm phys}\,+\,\mu^2\right) \vec{A}_g \;=\; - 16\,\pi\; G_N
\;\vec{J}_\varepsilon~. 
\label{eq:15}
\end{equation}
We see that the new term on the left-hand side, $(- 4
\dot{H}\;\vec{\beta})~{\rm resp}~(\mu^2\;\vec{A}_g)$, dominates for
superhorizon perturbations.  The solution is the Yukawa potential for
the source~$\vec{J}_\varepsilon$,
\begin{equation}
\vec{A}_g (\vec{x},t)\;=\;
\vec{\beta} \left(\vec{r},t\right) \;=\; -4\, G_N \int d^3\,
r'\,\vec{J}_\varepsilon ( \vec{r'}, t)~ \frac{{\rm
    exp}( - \mu |\vec{r}-\vec{r'}|)}{|\vec{r}-\vec{r'}|}~,
\label{eq:16}
\end{equation}
analogous to the formula for ordinary magnetostatics except for the
exponential cutoff. 
\end{enumerate}

\section{Mach's Principle}

The Yukawa potential in eq.~(\ref{eq:16}) has an exponential cutoff for
$|\vec{r}-\vec{r'}| \geq \frac 1\mu$, hence we reach our first
important conclusion: The contributions of vorticity perturbations
beyond the $\dot{H}$ radius are exponentially suppressed.

Our second result concerns the exact dragging of gyroscope axes by a
homogeneous rotation of cosmological matter out to significantly
beyond the $\dot{H}$ radius (for the exponential cutoff to be
effective). This holds for any equation of state. This is easily seen
from Einstein's $G_{\hat{0}\hat{\jmath}}$ equation in $k$-space for
superhorizon perturbations, $k_{\rm phys}^2 \;\ll\;(-\dot{H})$,
\begin{equation}
- 4 \dot{H} \vec{\beta} \;=\;- 16\pi\, G_N\;\vec{J}_\varepsilon,~~~
\vec{J}_\varepsilon \;=\; (\rho + p)\,\vec{v}_{\rm fluid},~~~
\dot{H} \;=\; - 4 \pi\;G_N (\rho+p)~.
\label{eq:17}
\end{equation}
All the prefactors cancel, and we obtain $\vec{\beta}(\vec{x}) = -
\vec{v}_{\rm fluid}(\vec{x}).$ With $\vec{\Omega}_{\rm gyroscope} = -
\frac{1}{2} \vec{B} = - \frac{1}{2}(\vec{\nabla} \times \vec{\beta})$
from eqs.~(\ref{eq:11}) and (\ref{eq:12}), and with $\vec{\Omega}_{\rm
  fluid} = \frac{1}{2} ( \vec{\nabla} \times \vec{v}_{\rm fluid})$ we
obtain
\begin{equation}
\vec{\Omega}_{\rm gyroscope} = \vec{\Omega}_{\rm fluid}.
\label{eq:18}
\end{equation}
This proves exact dragging of gyroscope axes here by a homogeneous
rotation of cosmological matter out to significantly beyond the
$\dot{H}$ radius. 

Our third result concerns the most general vorticity perturbation in
linear approximation, and it states what specific average of energy
flow in the universe determines the motion of gyroscope axes here at
$r=0$,
\begin{eqnarray}
&&\vec{\Omega}_{\rm gyroscope} (r=0,t)\,=\,-\frac 1 2\, \vec{B}_g
(r=0,t) \,=  \nonumber \\
&&~~~~~~~~2 G_N (\rho +p) \int d^3 
\, r\, \frac{1}{r^3}\, \left[\left( 1+\mu r\right) \,
  e^{-\mu\,r}\right] \left[ \vec{r} \times \vec{v} \left( \vec{r},
    t\right)\right]~. 
\label{eq:19}
\end{eqnarray}
The right-hand side of this equation is the gravitomagnetic moment of
the energy current distribution on a shell $[r,r+dr],$ analogous to
the magnetic moment of an electric current distribution, integrated
over $r$ with the given weight function.  This is the lowest term, the
$l=1$ term, in the multipole expansion for $r_{\rm obs}=0$ and $
r_{\rm source} > 0.$ Higher multipoles do not contribute to
$\vec{\Omega}_{\rm gyroscope}$ at $ r = 0.$ Only the toroidal velocity
field for $l=1$ of eq.~(\ref{eq:1}) contributes in eq.~(\ref{eq:19}),
i.e. a term corresponding to a rigid rotation velocity
$\vec{\Omega}(r)$ for each shell $(r,r+dr)$.

Eqs.~(\ref{eq:16}) and (\ref{eq:19}) are equations at fixed time, the
precession of a gyroscope axis here is determined by the rotational
motion of the masses in the universe at the same cosmic time.  This
shows that Einstein's objection to Mach's principle of 1949 \cite{3}
is not valid for weak gravitomagnetism. Furthermore, because of the
exponential cutoff, which follows from Einstein's equations in the
cosmological context, there is no need to impose ``an appropriate
boundary condition of some kind''.
%\end{enumerate}

Is exact dragging a measurable effect in principle? There seem to be
two problems:~ 1) To obtain exact dragging (within 
some given experimental error) the observational cutoff radius must be
significantly larger than the $\dot{H}$ radius (for the exponential
cutoff to be effective). We have defined $\vec{\Omega}_{\rm
  gyroscope}$ and $\vec{\Omega}_{\rm fluid}$ relative to distant
galaxies, i.e. galaxies beyond the rotational perturbation. But we
cannot ``see'' galaxies beyond the $\dot{H}$ radius. ~2) The
$G_{\hat{0}\hat{\jmath}}$ equation for ``star''-fixed FIDO's is an
equation at fixed time, our results hold at fixed time, but an
experimental test today with telescopes here on earth cannot see the
galaxies at the same cosmic time, it sees galaxies on the past light
cone.  The solution to these problems is an experimental test in
principle: Take data locally by many different observers all over
$\vec{x}$ space, also beyond the $\dot{H}$ radius.  Collect and patch
together the data after a few Hubble times.  This can be done as
a computer experiment, where the resulting  test only
involves our measurable quantities. This computer experiment is not
necessary, since we have given the proof (at the level of linear
perturbation theory).

We conclude that in  a spatially flat FRW Universe with rotational
perturbations (treated at the linear level) Mach's Principle on
nonrotating frames follows from General Relativity.

\Acknowledgements
We thank T. Cereghetti for collaboration while he was working on
his diploma thesis at ETH.

\end{document}